\documentclass[12pt,twocolumn]{aastex63}
\usepackage{bm}

\shorttitle{Magnetogenesis by charge exchange in an SNR}
\shortauthors{Kashiwamura and Ohira}
\usepackage{natbib}
\begin{document}

\title{Magnetic field generation by charge exchange in a supernova remnant in the early universe}

\author{Shuhei Kashiwamura}
\affiliation{Department of physics, The University of Tokyo, 7-3-1 Hongo, Bunkyo-ku, Tokyo 113-0033, Japan}
\email{kashiwamura-shuhei592@g.ecc.u-tokyo.ac.jp}

\author{Yutaka Ohira}
\affiliation{Department of Earth and Planetary Science, The University of Tokyo, 7-3-1 Hongo, Bunkyo-ku, Tokyo 113-0033, Japan}
\email{y.ohira@eps.s.u-tokyo.ac.jp}

\begin{abstract}
We present new generation mechanisms of magnetic fields in supernova remnant shocks propagating to partially ionized plasmas in the early universe. 
Upstream plasmas are dissipated at the collisionless shock, but hydrogen atoms are not dissipated because they do not interact with electromagnetic fields. 
After the hydrogen atoms are ionized in the shock downstream region, they become cold proton beams that induce the electron return current. 
The injection of the beam protons can be interpreted as an external force acting on the downstream proton plasma. 
We show that the effective external force and the electron return current can generate magnetic fields without any seed magnetic fields. 
The magnetic field strength is estimated to be $B\sim 10^{-14}-10^{-11}~{\rm G}$, where the characteristic lengthscale is the mean free path of charge exchange, $\sim 10^{15}~{\rm cm}$. 
Since protons are marginally magnetized by the generated magnetic field in the downstream region, the magnetic field could be amplified to larger values and stretched to larger scales by turbulent dynamo and expansion. 
\end{abstract}

\keywords{Cosmic magnetic fields theory (321), Astrophysical magnetism (102), Magnetic fields (994), Galaxy magnetic fields (604), Plasma astrophysics (1261), Plasma physics (2089)}

\section{Introduction}
Magnetic fields are ubiquitous and play an important role in the current universe, from planet to intergalactic scales \citep{han17,akahori18}. 
The Earth and our lives are protected from the solar wind, solar energetic particles, and cosmic rays by the Earth's magnetic field. 
The magnetic field contributes to the angular momentum transfer in accreting systems \citep{balbus91} and acceleration of nonthermal high energy particles \citep{axford77,krymsky77,bell78,blandford78}. 
In addition, the magnetic field can transfer or extract energy from some sources, e.g. black-hole spin and kinetic energy of plasma motions, so that plasmas connected with the energy sources by the magnetic field line are heated or accelerated \citep{blandford82,suzuki02,fujita07,fujita11}.

Although various mechanisms of the magnetic field generation in the early universe have been considered, 
the origin of magnetic fields is still an open issue \citep{widrow02}. 
We do not understand which mechanism generates which scale's magnetic field in the current universe.  
For magnetic field generations by astrophysical mechanisms, the electron pressure \citep{biermann50}, the electron ram pressure induced by the cosmic-ray streaming \citep{ohira20}, the radiation force acting on electrons \citep{harrison70}, and the friction force acting on electrons \citep{lesch89} have been proposed as a driving force to generate the electric current. These forces act on only electrons.  
In this work, we propose a new generation mechanism of magnetic fields, which does not need a force acting on electrons.

Most baryonic matters are neutral hydrogen atoms in the early universe before the reionization epoch, and in  protogalaxies even after the reionization epoch. 
\citet{lesch89} and \citet{huba93} proposed that interactions between free electrons and hydrogen atoms generate magnetic fields in a weakly ionized plasma with a shear flow. 
In those work, interactions between protons and hydrogen atoms were not taken into account when the generalized Ohm's law was derived. 
However, this treatment is not always correct. 
In this work, we more correctly derive the generalized Ohm's law in a partially ionized plasma.

Many massive stars explode as supernovae in protogalaxies. 
\citet{hanayama05} considered a magnetic field generation by the Biermann battery in the shock downstream region of supernova remnants (SNRs). 
They showed by hydrodynamical simulations that magnetic fields with the strength of $B\sim 10^{-17} - 10^{-14}{\rm G}$ are generated. 
The Biermann battery needs inhomogeneous density and pressure distributions.
However, the inhomogeneity is expected to decay by particle diffusion during the sound crossing timescale 
because the downstream plasma of SNRs is collisionless. 
If the diffusion is efficient, the Biermann battery cannot generate the magnetic field efficiently, but  another magnetic field generation by diffusion could work \citep{tomita16}. 
Even for no large scale magnetic field, cosmic rays can be accelerated by SNRs in the early universe because collisionless shocks generate magnetic fields in a small scale \citep{ohira19}. 
This small scale magnetic field would rapidly decay in the shock downstream region after the upstream flow energy is fully dissipated at the shock front. 
The cosmic rays in the early universe can generate large scale magnetic fields by some mechanisms \citep{miniati11,ohira20,ohira21}

In this work, we propose a new mechanism of the magnetic field generation at SNR shocks propagating to partially ionized plasmas. 
For collisionless shocks in partially ionized plasmas, upstream hydrogen atoms freely penetrate the shock front without dissipation. 
Then, the penetrating hydrogen atoms are ionized mainly by the charge exchange process in the downstream region. 
After the charge exchange reaction, the penetrating cold hydrogens become cold proton beams, which can be interpreted as the momentum injection or a force acting on the downstream proton plasma. 
The effective force acting on protons drives the electric current, which is a remarkable difference from work on the magnetic field generation so far. 
In addition to the magnetic field generation by the effective force acting on protons, the magnetic field can be generated by the electron return current induced by the proton beam \citep{ohira20,ohira21}. 
In Section \ref{sec2}, we first briefly review the collisionless shock in a partially ionized plasma. 
Then, we derive the generalized Ohm's law in a partially ionized plasma in Section \ref{sec3}. 
In Section \ref{sec4}, we estimate how strong magnetic fields are generated by the new mechanism and the other mechanisms. 
Sections \ref{sec5} and \ref{sec6} are devoted to the discussion and summary, respectively.

\section{Collisionless shock in a partially ionized plasma}
\label{sec2}
In the current universe, many SNRs are expanding to partially ionized plasmas \citep{heng10}. 
H$\alpha$ emission is actually observed from SNRs \citep{chevalier78}. 
Therefore, SNR shocks in protogalaxies in the early universe are expected to propagate to partially ionized plasmas. 
In this section, we briefly review the collisionless shock structure in a partially ionized plasma with a small neutral fraction. 
The collisionless shock structure has not been understood in weakly ionized plasmas. 
We here adopt the shock rest frame.

For plasma components, the upstream flow velocity, $u_1$, is decelerated to the downstream flow velocity, $u_2=u_1/r$ by electromagnetic interactions around the collisionless shock front. 

For a high Mach number shock with the adiabatic index of $5/3$, the compression ration, $r$, becomes almost four. 
 
Then, the temperature and density of the downstream plasma become large compared with those in the upstream region. 
SNR shocks are magnetized collisionless shocks in the current universe as long as the shock velocity is faster than about $100~{\rm km/s}$ \citep{ohira12}. 
In the early universe, small scale magnetic fields are generated by the 

Weibel or current filamentation instability \citep{weibel59,bret10}

in the shock transition region, which disturb the upstream plasma flow \citep{ohira19}. 
Either way, the lengthscale of the collisionless shock transition region is on the order of the ion gyroradius, which is much smaller than the mean free path of charge exchange. 
On the other hand, upstream cold hydrogen atoms freely penetrate the  collisionless shock front without dissipation and deceleration because they do not interact with electromagnetic fields. 

In the shock downstream region, the penetrating cold hydrogen is ionized by the charge exchange process with hot protons and collisional ionization with hot electrons and protons. 
Since the rate of charge exchange is roughly a few times larger than one of collisional ionization with hot electrons and protons as long as the shock velocity is smaller than about $3000~{\rm km/s}$ \citep{heng07}, 
the charge exchange process is the dominant ionization process in the typical SNRs. 
On the other hand, for faster shocks, the hydrogen atoms are mainly ionized by the collisional ionization with hot electrons and protons because the cross section of charge exchange has a cutoff signature at the relative velocity of $3000~{\rm km/s}$. 

In the charge exchange process, the upstream hydrogen becomes proton while the downstream proton becomes hydrogen. 
This picture was first proposed by \citet{chevalier78} and confirmed by kinetic plasma simulations \citep{ohira13}.

After the ionization, the penetrating cold hydrogen atoms become cold beam protons in the downstream region. 
Therefore, there are two proton populations in the downstream region, 
hot protons that experienced the shock dissipation and cold protons that originate from the penetrating hydrogen atoms. 
The mean velocities of hot and cold protons are $u_2$ and $u_1$, respectively. 
In the current universe, this proton distribution in the velocity space excites many kinetic plasma instabilities \citep{raymond08,ohira09}, so that the proton distribution is expected to rapidly relax to a single population. 
In the early universe, if there is no large scale magnetic field, the cold proton beam would excite the 
Weibel or current filamentation instability,
but it is unclear how long it takes for the cold protons to relax to the hot protons. 

The Weibel or current filamentation instability generate small scale magnetic fields whose lengthscale is initially about the electron or proton inertial lengthscales. 
\citet{kehset09} showed by a long-term particle-in-cell simulation that the characteristic length scale of magnetic fields generated by the Weibel or current filamentation instability increases with time, so that the decay timescale of the magnetic field becomes longer in the shock downstream region. 
Their simulation timescale is about $10^4\ \omega_{\rm pe}^{-1}$, where $\omega_{\rm pe}$ is the plasma frequency. 
For nonrelativistic shocks in the electron density of $1~{\rm cm}^{-3}$, their timescale of $10^4\ \omega_{\rm pe}^{-1}$ is $\approx 7~ {\rm sec}$ that is much smaller than the astrophysical timescale. 
Therefore, whether it decays or grows to the large scale magnetic field for the astrophysical timescale is still an open question \citep{tomita16,tomita19}.

The downstream hot protons become hot hydrogen atoms by charge exchange. 
The velocity dispersion of the hot hydrogen is comparable to the downstream flow velocity, $u_2$. 
In addition, the hot hydrogen atoms can freely penetrate the shock front, so that some of the hot hydrogen atoms can leak to the upstream region \citep{lim96,blasi12,ohira12}. 
The leaking hydrogen atoms are ionized in the upstream region by charge exchange or collisional ionization with the upstream matter. 
Then, the leaking hydrogen atoms become a hot proton beam in the upstream region.
For SNRs in the current universe, 
some plasma instabilities are excited and the shock structure is modified \citep{ohira13,ohira14,ohira16a,ohira16b}.
In the early universe, if there is no large scale magnetic field, the hot proton beams would excite the 
Weibel or current filamentation
instability like the cold proton beam in the downstream region.

As mentioned above, when a collisionless shock propagates to a partially ionized plasma, beam protons are injected by ionization of hydrogen atoms both in the upstream and downstream regions. 
The injection of beam protons means that there is a momentum injection in the proton plasma component. 
In the next section, we show that the momentum injection induces the electric field. 
If the curl of the electric field is nonzero, the magnetic field is generated. 
Furthermore, the beam protons induced the electron return current that can generate the magnetic field by new types of the Biermann battery \citep{ohira20,ohira21}

\section{Generalized Ohm's law}
\label{sec3}

In this section, we derive the generalized Ohm's law around a collisionless shock in a partially ionized plasma, where beam protons are injected by ionization of penetrating or leaking hydrogen atoms. 
In this work, for simplicity, we consider electrons, protons, beam protons, and hydrogen atoms as particles. 

All components generally have different drift (or bulk) velocities, $\bm{V}_{\rm e}, \bm{V}_{\rm p}, \bm{V}_{\rm b}$, and $\bm{V}_{\rm H}$. 

Furthermore, we consider only charge exchange as ionization of hydrogen atoms, which is valid as long as the shock velocity is smaller than $3000~{\rm km/s}$ \citep{heng07}.
Equations of motion for each plasma component are as follows: 
\begin{eqnarray}
\frac{\partial}{\partial t}(m_s n_s \bm{V}_s)&+&\bm{\nabla} \cdot (m_s n_s \bm{V}_s \bm{V}_s)=  \nonumber \\
&-&\bm{\nabla} p_s  + q_s n_s \left(\bm{E}+\frac{\bm{V}_s\times \bm{B}}{c}\right)  \nonumber \\
&-& n_s m_s \sum_{s'\neq s} \frac{(\bm{V}_s-\bm{V}_{s'})}{\tau_{ss'}} -\bm{\nabla} \pi_s \nonumber \\
&+&Q_{{\rm mom},s} ~~,
\label{eq:beam}
\end{eqnarray}
where $m_s ,n_s, {\bm V}_s, p_s, q_s, {\bm E}, {\bm B}, c$, {\bf and $\pi_s$} are the mass, number density, velocity, pressure, charge, electric field, magnetic field, speed of light, and viscous stress tensor, respectively. 
The subscripts $s(={\rm e,p,b})$ and $s'(={\rm e,p,b,H})$ denote the particle species, where e, p, b, and H mean electrons, protons, beam protons, and hydrogen atoms, respectively. 

In this work, an isotropic momentum distribution is assumed for each plasma in each plasma rest frame as a simple example, so that the pressures are not tensor but scalar. 

The terms including $\tau_{s s'}$ represent the momentum transfer between particle $s$ and particle $s'$ by collision, where $\tau_{ss'}$ is the timescale of momentum transfer due to collision. 
The last term on the right-hand side, $Q_{{\rm mom},s}$, represents the momentum injection due to charge exchange.

After a charge exchange process happens, a hydrogen atom becomes a beam proton with the momentum of $m_{\rm p}V_{\rm H}$, a proton with the momentum of $m_{\rm p}V_{\rm p}$ becomes hydrogen atoms, and no electron is newly generated. Therefore, the momentum injection terms are given by
\begin{eqnarray}
Q_{\rm mom,e} &=& 0 ~~,\\
Q_{\rm mom,p} &=& - m_{\rm p} n_{\rm H} \frac{\bm{V}_{\rm p}}{\tau_{\rm CE}},\\
Q_{\rm mom,b} &=&  m_{\rm p} n_{\rm H} \frac{\bm{V}_{\rm H}}{\tau_{\rm CE}},
\label{eq:qion}
\end{eqnarray}
The timescale of charge exchange for each hydrogen atom, $\tau_{\rm CE}$, is given by 
\begin{eqnarray}
\tau_{\rm CE}&=&(n_{\rm p}\sigma_{\rm CE}v_{\rm rel})^{-1}~~,\nonumber \\
&=& 10^{7}\ {\rm sec} \left(\frac{n_{\rm p}}{1 {\rm cm}^{-3}}\right)^{-1} \left(\frac{\sigma_{\rm CE}}{10^{15} {\rm cm}^2}\right)^{-1}  \left(\frac{v_{\rm rel}}{10^8 {\rm cm/s}}\right)^{-1},
\end{eqnarray}
where $\sigma_{\rm CE}\sim 10^{-15}~{\rm cm}^2$ is the cross section of charge exchange \citep{schultz08} and $v_{\rm rel}$ is the mean relative velocity between protons and hydrogen atoms. 

Multiplying Equation~(\ref{eq:beam}) by $q_s / m_s$ and summing these equations for charged particles, we obtain 
\begin{eqnarray}
\frac{\partial \bm{J}}{\partial t}&+&\bm{\nabla} \cdot \left(\sum_s q_s n_s \bm{V}_s \bm{V}_s \right) = \nonumber \\
&-& \sum_s \frac{q_s}{m_s}\bm{\nabla} p_s+\sum_s\frac{q_s^2 n_s}{m_s} \left(\bm E+\frac{\bm V_s \times \bm B}{c}\right) \nonumber \\ 
&-& \sum_s \sum_{s'\neq s} \frac{q_s n_s (\bm{V}_s -\bm{V}_{s'})}{\tau_{ss'}}  -\sum_s \frac{q_s}{m_s} \bm{\nabla} \pi_s \nonumber \\ 
&+& \frac{en_{\rm H}(\bm{V}_{\rm H}-\bm{V}_{\rm p})}{\tau_{\rm CE}}
\label{eq:sum}
\end{eqnarray}
where $\bm{J} \equiv \sum_s q_s n_s \bm{V}_s $ is the electric current. 
 
The above equation describes the time evolution of the total current.

One can derive the generalized Ohm's law by solving this equation for the electric field.

The first term on the left-hand side of the above equation is negligible if one considers a large scale compared with the electron inertial lengthscale. 
The second term on the left-hand side has an important role in the magnetic field generation if plasmas consist of more than two components \citep{ohira20}. 
In the first, 
second, third and fifth
terms on the right-hand side, the electron contribution dominates the others because of the large mass ratio, $m_{\rm p}/m_{\rm e}\gg1$.

The third term on the right-hand side is negligible as long as the magnetic field is sufficiently weak. Even though the third term becomes significant in the generalized Ohm's law, the third therm does not generate magnetic fields. Although the third term can amplify the generated magnetic fields by magnetohydrodynamics, we are interested in the generation of magnetic fields in this work. 
Therefore, we ignore the third term in this work. 

Then, the generalized Ohm's law is reduced to 
\begin{eqnarray}
\bm{E} &=& -\frac{\bm{\nabla} p_{\rm e} }{en_{\rm e}}+ \frac{m_{\rm e}}{e^2 n_{\rm e}}  \bm{\nabla} \cdot \left(\sum_s q_s n_s \bm{V}_s \bm{V}_s \right)  \nonumber \\
          &+& \frac{m_{\rm e}}{e^2 n_{\rm e}} \sum_s \sum_{s'\neq s} \frac{q_s n_s (\bm{V}_s -\bm{V}_{s'})}{\tau_{ss'}} -\frac{\bm{\nabla} \pi_{\rm e} }{en_{\rm e}} \nonumber \\
          &-& \frac{m_{\rm e}n_{\rm H}(\bm{V}_{\rm H}-\bm{V}_{\rm p})}{en_{\rm e}\tau_{\rm CE}} ~~.
\label{eq:ohm1}
\end{eqnarray}

The first and second terms on the right-hand side describe the Biermann battery by the thermal and ram pressures, respectively \citep{biermann50,ohira20,ohira21}. 
If the background plasma and hydrogen have no velocity fluctuations, only the electron flow has inhomogeneity because a nonuniform density distribution makes the electron return current nonuniform. 
Since the magnetic field generation needs the curl of the electric field, in the second term, only the electron momentum flux contributes the magnetic field generation.  
Since the third term originates from the momentum transfer between two components, it describes the resistivity that usually dissipates the magnetic field. 
However, there are some interesting cases where the magnetic field is generated by the resistive term \citep{bell03,miniati11,lesch89,huba93}. 
\citet{lesch89} considered a friction between an expanding plasma and a rotating neutral gas in nuclear regions of galaxies, 
where the relative velocity between the plasma and the hydrogen atoms is on the order of a few $100~{\rm km/s}$. 
Then, electrons are scattered more frequently than protons by rotating hydrogen atoms, so that an electric ring current is generated. 
In this work, we consider a collsionless shock with the shock velocity of a few $1000~{\rm km/s}$. 

Then, all the timescales of the Coulomb collision and the timescale of electron-hydrogen collision become longer than the timescale that we consider in this work, $\tau_{\rm CE} \approx 10^7\ {\rm sec}$, so that the plasma system is still collisionless plasma \citep{spitzer62,pinto08,schultz08}. Therefore, the third and fourth terms originating from collisions cannot play an important role in this collisionless plasma system. The fifth
term originates from the momentum injection due to charge exchange. 
Therefore, if the curl of the forth term is nonzero, the magnetic field is generated, which has not been considered so for.  

By adopting the proton rest frame ($\bm{V}_{\rm p}=0$), the generalized Ohm's law is reduced to  
\begin{equation}
\bm{E} = -\frac{{\bm \nabla} p_{\rm e}}{en_{\rm e}} -\frac{m_{\rm e}}{e^2 n_{\rm e}} \bm{\nabla} \cdot \left( \sum_{s={\rm e,b}} q_s n_s \bm{V}_s \bm{V}_s\right) -  \frac{m_{\rm e}}{e} \frac{n_{\rm H}}{n_{\rm e}}\frac{\bm{V}_{\rm H}}{\tau_{\rm CE}} ~,
\label{eq:ohm2}
\end{equation}
where the velocity field of protons is assumed to be uniform. 
In the shock system, the proton rest frame corresponds to the upstream or downstream rest frame.

\section{Magnetic field generation}
\label{sec4}
The curl of the electric field must be nonzero to generate the magnetic field by the Faraday's law. 
It has been shown by previous work that the first two terms on the right-hand side of Equation~(\ref{eq:ohm2}) can have a finite curl \citep[e.g.][]{ohira20,ohira21}. 
The curl of the third term in Equation~(\ref{eq:ohm2}) is proportional to $(n_{\rm H}/n_{\rm e}\tau_{\rm CE}) \bm{\nabla}\times \bm{V}_{\rm H} - \bm{V}_{\rm H} \times \bm{\nabla}(n_{\rm H}/n_{\rm e}\tau_{\rm CE})$. 
The direction of the hydrogen velocity is parallel or antiparallel to the shock normal direction. 
Hence, if the direction of the density gradient is not parallel to the shock normal direction, the curl of the electric field generated by charge exchange is nonzero, so that it can generate the magnetic field.

From the Faraday's equation, the magnetic field strength can be estimated by $B \sim E (ct/L)$, 
where $E, t$, and $L$ are the electric field strength, characteristic timescale and lengthscale. 
As one can see in Equation~(\ref{eq:ohm2}), the electric field consists of three terms. 
In the following subsection, we estimate the order of magnitude of magnetic fields generated by each term in Equation~(\ref{eq:ohm2}).

\subsection{Magnetic field generation by charge exchange}
We first estimate the magnetic field strength generated by charge exchange. 
In an SNR shock propagating to a partially ionized plasma, 
leaking hydrogen atoms in the upstream region and penetrating hydrogen atoms in the downstream region become beam protons by the charge exchange process. 
The momentum injection of beam protons can be interpreted as a force acting only on protons, so that the electric current is generated. 
This magnetic field generation mechanism has not been considered so for as far as we know. 
From the third term in Equation~(\ref{eq:ohm2}), the magnetic field strength is estimated to be 
\begin{equation}
B_{\rm CE} \sim \frac{m_{\rm e}cn_{\rm H}V_{\rm H}}{en_{\rm e}L}~~,
\label{eq:bce1}
\end{equation}
where $t=\tau_{\rm CH}$ is used because after all hydrogen atoms are ionized, this mechanism does not work.
In this mechanism, the maximum lengthscale of the magnetic field is the mean free path of the charge exchange process for hydrogen, 
\begin{eqnarray}
l_{\rm CE} &=& (n_{\rm p}\sigma_{\rm CE})^{-1}\nonumber \\
&\sim& 10^{15} {\rm cm}~\left(\frac{\sigma_{\rm CE}}{10^{-15} {\rm cm}^2}\right)^{-1} \left(\frac{n_{\rm p}}{1~{\rm cm}^{-3}}\right)^{-1}~~.
\label{eq:lce}
\end{eqnarray}
Then, the magnetic field strength becomes 
\begin{eqnarray}
B_{\rm CE,m} &\sim& \frac{m_{\rm e}c \sigma_{\rm CE} n_{\rm H} V_{\rm H}}{e} \nonumber \\
&\sim& 5.7 \times 10^{-15}~{\rm G} \left( \frac{n_{\rm H}}{1~{\rm cm}^{-3}}\right) \left( \frac{V_{\rm H}}{10^8~{\rm cm/s}}\right)
\label{eq:bce2}
\end{eqnarray}
For a smaller lengthscale ($L<l_{\rm CE}$), the magnetic field strength becomes larger than the above value ($B\propto L^{-1}$). 

\subsection{Magnetic field generation by the momentum flux}
Next, we estimate the magnetic field strength generated by the momentum flux.
After the ionization of leaking or penetrating hydrogen atoms, the beam protons induces the electron return current in order to neutralize the current and charge of the beam protons. 
If the electron or beam-proton densities have some inhomogeneities, the electron return current also has some inhomogeneities, so that the magnetic field is generated \citep{ohira20}. 

In the proton rest frame corresponding to the upstream or downstream rest frame in the shock system, the electron velocity induced by the return current is obtained from the current neutrality condition, $-en_{\rm e}\bm{V}_{\rm e} + en_{\rm b} \bm V_{\rm H}=\bm{0}$. 
\begin{eqnarray}
\bm{V}_{\rm e} &=& \frac{n_{\rm b}}{n_{\rm e}} \bm V_{\rm H} \nonumber \\
&\approx&  \frac{n_{\rm H}}{n_{\rm e}} \bm{V}_{\rm H} \times \left\{ \begin{array}{ll}
\left(t/\tau_{\rm CE}\right)                     & ~(~t \leq \tau_{\rm CE}~) \\
1 & ~(~t \geq \tau_{\rm CE}~) \\
\end{array} \right. ~~,
\label{eq:ve}
\end{eqnarray}
where $n_{\rm H}$ is the number density of hydrogen in the upstream region. 
We assumed that the beam protons leave their drift velocity $V_{\rm H}$ even after all hydrogens are ionized. 
From Equation (\ref{eq:ve}) and the second term of Equation (\ref{eq:ohm2}), the magnetic field strength is estimated to be
 \begin{eqnarray}
 B_{\rm MF} \sim \frac{m_{\rm e}cn_{\rm H}^2V_{\rm H}^2\tau_{\rm CE}}{en_{\rm e}^2L^2} \times \left\{ \begin{array}{ll}
\left(t/\tau_{\rm CE}\right)^3 & ~(~t \leq \tau_{\rm CE}~) \\
\left(t/\tau_{\rm CE}\right) & ~(~t \geq \tau_{\rm CE}~) \\
\end{array} \right. ~~.
\label{bmf}
\end{eqnarray}
At $t=\tau_{\rm CE}$ with $L=l_{\rm CE}=V_{\rm H}\tau_{\rm CE}$, the magnetic field strength becomes $B_{\rm MF,m}\sim B_{\rm CE,m} (n_{\rm H}/n_{\rm e})$.

Even after the charge exchange timescale ($t>\tau_{\rm CE}$), the inhomogeneous return current can further generate magnetic fields if the beam protons have a relative velocity with respect to the background protons. 
In this case, the longest timescale for the magnetic field generation is expected to be $L/V_{\rm e}$, so that the magnetic field strength becomes $B_{\rm MF} \sim B_{\rm CE}$ (see Equation (\ref{eq:bce1}) for $B_{\rm CE}$). 
Therefore, the electron return current induced by the beam protons and the charge exchange process generate the same level of magnetic fields.

\subsection{Magnetic field generation by the electron pressure}
We finally estimate the magnetic field strength generated by the electron pressure, that is, the Biermann battery. 
In order for the Biermann battery to work, the electron pressure needs the special profile that stratifies the condition of ${\bm \nabla} n_{\rm e} \times {\bm \nabla} p_{\rm e} \neq \bf{0}$. 
\citet{ohira21} showed that the electron flow induced by the electron return current can generate the special profile. The electron pressure gradient is estimated to be 
\begin{equation}
\nabla p_{\rm e} \sim \frac{\Gamma t  V_{\rm e} p_{\rm e}}{L^2} ~~,
\label{eq:pe}
\end{equation}
where $\Gamma$ is the adiabatic index \citep{ohira21}. 
From Equations (\ref{eq:ve}), (\ref{eq:pe}), and the first term of Equation (\ref{eq:ohm2}), the magnetic field strength is estimated to be
 \begin{eqnarray}
 B_{\rm EP} \sim \frac{c\Gamma p_{\rm e}n_{\rm H}V_{\rm H}\tau_{\rm CE}^2}{en_{\rm e}^2L^3} \times \left\{ \begin{array}{ll}
\left(t/\tau_{\rm CE}\right)^3 & ~(~t \leq \tau_{\rm CE}~) \\
\left(t/\tau_{\rm CE}\right)^2 & ~(~t \geq \tau_{\rm CE}~) \\
\end{array} \right. ~~.
\label{bep}
\end{eqnarray}
 At $t=\tau_{\rm CE}$ with $L=l_{\rm CE}=V_{\rm H}\tau_{\rm CE}$, the magnetic field strength becomes $B_{\rm EP,m}\sim B_{\rm CE,m} (c_{\rm s,e}/V_{\rm H})^2$, where $c_{\rm c,s}=(\Gamma p_{\rm e}/m_{\rm e}n_{\rm e})^{1/2}$ is the sound velocity of electrons. 
Compared with the upstream region, the electron sound velocity becomes large in the shock downstream region, so that the Biermann battery driven by the electron return current generates magnetic field efficiently in the downstream region. 
For SNRs in the current universe, the electron temperature in the shock downstream region is estimated to be $T_{\rm e}/ T_{\rm p} \sim 10^{-2} - 10^{-1}$ \citep{ghavamian07,adelsberg08,ohira07,ohira08}. 
From shock conditions for the high Mach number limit ($T_{\rm p}=3m_{\rm p}V_{\rm sh}/16$ and $V_{\rm H}=3V_{\rm sh}/4$), one can obtain $(c_{\rm s,e}/V_{\rm H})^2\approx 10^3 (T_{\rm e}/ T_{\rm p})$, where $V_{\rm sh}$ is the shock velocity. 
By adopting the temperature ratio of $T_{\rm e}/ T_{\rm p} \sim 10^{-1}$, the magnetic field strength at $t=\tau_{\rm CE}$ with $L=l_{\rm CE}$ becomes
\begin{eqnarray}
 B_{\rm EP,m} &\sim& 10^2 B_{\rm CE} \left(\frac{T_{\rm e}/T_{\rm p}}{10^{-1}}\right) \nonumber \\
  &\sim& 5.7 \times 10^{-13}~ {\rm G} \left( \frac{n_{\rm H}}{1~{\rm cm}^{-3}}\right) \left( \frac{V_{\rm H}}{10^8~{\rm cm/s}}\right) \left(\frac{T_{\rm e}/T_{\rm p}}{10^{-1}}\right)  ~~.
\end{eqnarray}
For a smaller lengthscale ($L<l_{\rm CE}$), the magnetic field strength at $t=\tau_{\rm CE}$ becomes larger than the above value ($B\propto L^{-3}$).

Even after the charge exchange timescale ($t>\tau_{\rm CE}$), the electron pressure can further generates magnetic fields as well as the momentum flux does. 
At $t=L/V_{\rm e}$, the magnetic field strength becomes 
\begin{eqnarray}
B_{\rm EP} &\sim& B_{\rm CE}  \left(\frac{c_{\rm s,e}}{V_{\rm H}}\right)^2 \left(\frac{n_{\rm e}}{n_{\rm H}}\right)^2 \nonumber \\
                   &\sim&\frac{ c\Gamma p_{\rm e}}{en_{\rm H}V_{\rm H}L}~~. 
\end{eqnarray}
If the lengthscale is the charge exchange scale, $L=l_{\rm CE}$, the magnetic field strength at $t=l_{\rm CE}/V_{\rm e}=\tau_{\rm CE}(n_{\rm e}/n_{\rm H})$ is 
\begin{eqnarray}
B_{\rm EP} &\sim& B_{\rm CE,m}  \left(\frac{c_{\rm s,e}}{V_{\rm H}}\right)^2 \left(\frac{n_{\rm e}}{n_{\rm H}}\right)^2 \nonumber \\
                   &=&5.7 \times 10^{-12}~ {\rm G} \left( \frac{n_{\rm e}}{1~{\rm cm}^{-3}}\right)\left( \frac{n_{\rm e}/n_{\rm H}}{10}\right) \nonumber \\
                   &&~~~~~~~~~~~~~~~~\times \left( \frac{V_{\rm H}}{10^8~{\rm cm/s}}\right) \left(\frac{T_{\rm e}/T_{\rm p}}{10^{-1}}\right)~~. 
\label{eq:bepmax}
\end{eqnarray}
For a lower neutral fraction, $n_{\rm e}/n_{\rm H}\gg 1$, the generated magnetic field becomes larger because the generation time, $t=\tau_{\rm CE}(n_{\rm e}/n_{\rm H})$, becomes longer. 
Therefore, compared with other two mechanisms, the Biermann battery driven by the electron return current generates strong magnetic fields in the shock downstream region.

\section{Discussion}
\label{sec5}
We assumed that beam protons produced by charge exchange preserve their drift velocity. 
Although the Weibel or current filamentation instability instability would be driven by the beam protons, 
whether the drift velocity of the beam protons is completely dissipated by the kinetic instability or not, and how long does it take to be dissipated have not been investigated. 
If the drift velocity of the beam protons is rapidly dissipated, magnetic fields are generated only by the momentum injection due to charge exchange (the third term on the right hand side of Equation (\ref{eq:ohm2})) because no electron return current is induced. 
Moreover, the beam protons are isotropized when the gyroradius of beam protons becomes comparable to the characteristic lengthscale of the generated magnetic field. From the magnetization condition,  $cm_{\rm p}V_{\rm H}/eB_{\rm c} \sim L$, the critical magnetic field strength is given by
\begin{eqnarray}
B_{\rm c} &\sim& \frac{m_{\rm p} c V_{\rm H}}{eL} \nonumber \\
&\sim& 10^{-11}~{\rm G} \left( \frac{L}{10^{15}~{\rm cm}} \right)^{-1} \left( \frac{V_{\rm H}}{10^8~{\rm cm/s}} \right) ~~, 
\label{eq:bs}
\end{eqnarray}
which is comparable to the estimation in Equation (\ref{eq:bepmax}).

In this work, the characteristic lengthscale is the mean free path of the charge exchange process for hydrogen (Equation (\ref{eq:lce})), which is much smaller than the astrophysical scale. 
Thanks to the small scale, the magnetic field is rapidly generated compared with other astrophysical mechanisms that directly generate magnetic fields with the astrophysical scale. 
Hence, the turbulent dynamo in SNRs or protogalaxies would quickly amplify the magnetic field generated by mechanisms proposed in this work. 
In order for the turbulent dynamo to work efficiently, the gyroradius of thermal protons should be comparable to or smaller than the characteristic lengthscale of the magnetic field \citep{rincon16,pusztai20}. 
As discussed in the previous paragraph, the magnetization condition could be marginally satisfied because the thermal velocity of downstream protons is comparable to the hydrogen velocity, $V_{\rm H}$. 
Even for unmagnetized collisionless shocks, some particles are accelerated to higher energies \citep{ohira19}. The accelerated particles have the larger gyroradius and could generate magnetic fields with the larger lengthscale \citep{ohira20,ohira21}.

In this work, it was assumed that the thickness of the shock transition region of plasmas is much smaller than the charge exchange lengthscale. 
If the neutral fraction is large, protons generated by ionization of hydrogen atoms modify the shock structure of plasmas, so that the thickness of the shock transition region becomes comparable to the charge exchange lengthscale. 
Most of the upstream kinetic energy is dissipated by electromagnetic interactions after the upstream hydrogen atoms are ionized. 
Hence, the Weibel or current filamentation instability or other magnetic field generation process would work in the dissipation region. 
Similar to the ionization front \citep{subramanian94,gnedin00}, the shock front has strong gradients of electron density and pressure, so that the Biermann battery could generate the magnetic field if the shock propagates a nonuniform medium. 
The pressure gradient of electrons is estimated to be $\nabla p_{\rm e} \sim (T_{\rm e}/{T_{\rm p}})m_{\rm p}n_{\rm p}V_{\rm sh}^2/l_{\rm CE}$, so that the magnetic field strength is estimated to be 

\begin{eqnarray}
B &\sim& \frac{T_{\rm e}m_{\rm p}cV_{\rm sh}}{T_{\rm p}el_{\rm CE}} \nonumber \\
&\sim& 10^{-12}~{\rm G} \left(\frac{T_{\rm e}/T_{\rm p}}{10^{-1}}\right) \left( \frac{n}{1~{\rm cm^3}} \right) \left( \frac{V_{\rm sh}}{10^8~{\rm cm/s}} \right) ~~, 
\label{eq:bs}
\end{eqnarray}
where the generation timescale is assumed to be $l_{\rm CE}/V_{\rm sh}$.
There is no study about the collisionless shock structure in the weakly ionized plasma, 
which should be addressed in future.

\section{Summary}
\label{sec6}
In this paper, we have investigated the magnetic field generation by the charge exchange process in SNR shocks propagating to a partially ionized plasma in the early universe. 
Upstream hydrogen atoms are not dissipated at the collisionless shock front and ionized in the shock downstream region. 
After the ionization, the cold hydrogen atoms become beam protons that induce the electron return current. 
We have derived the generalized Ohm's law that includes effects of the proton momentum injection due to the ionization, the electron momentum flux due to the return current, and the electron pressure (see Equations~(\ref{eq:ohm1}) and (\ref{eq:ohm2})). 
We have estimated the magnetic field strength generated by three mechanisms, which is $B\sim 10^{-14}-10^{-11}~{\rm G}$ (see Section~\ref{sec4}). 
The characteristic lengthscale is the mean free path of charge exchange, $\sim 10^{15}~{\rm cm}$, that is much smaller than the astrophysical lengthscale. 
Nevertheless, since protons would be marginally magnetized by the generated magnetic field in the shock downstream region, the turbulent dynamo could amplify the magnetic fields. 
Expansion of SNRs and winds from protogalaxies would make the characteristic lengthscale astrophysical scales. 

\acknowledgments
We thank the anonymous referee for his/her valuable comments that improved the paper.
This work is supported by JSPS KAKENHI Grant Number JP19H01893, and by Leading Initiative for Excellent Young Researchers, MEXT, Japan.


\begin{thebibliography}{}

\bibitem[Akahori et al.(2018)]{akahori18}
Akahori, T. et al. 2018, \pasj, 70, R2

\bibitem[Axford et al.(1977)]{axford77}
Axford, W. I., Leer, E., \& Skadron, G. 1977, Proc. 15th Int. Cosmic Ray Conf., (Plovdiv: Bulgarian Academy of Sciences), 11, 132

\bibitem[Bell(1978)]{bell78}
Bell, A. R. 1978, \mnras, 182, 147

\bibitem[Bell \& Kingham(2003)]{bell03}
Bell, A. R., \& Kingham, R. J., 2003, \prl, 91, 035003

\bibitem[Biermann(1950)]{biermann50}
Biermann, L. 1950, Zeitschrift f\"{u}r Naturforschung A, 5, 65

\bibitem[Balbus \& Hawley(1991)]{balbus91}
Balbus, S. A., \& Hawley. J. F. 1991, \apj, 376, 214

\bibitem[Blandford \& Ostriker(1978)]{blandford78}
Blandford, R. D., \& Ostriker. J. P. 1978, \apj, 221, 29

\bibitem[Blandford \& Payne(1982)]{blandford82}
Blandford, R. D., \& Payne. D. G. 1982, \mnras, 199, 883

\bibitem[Blasi et al.(2012)]{blasi12}
Blasi, P., Morlino, G., Bandiera, R., Amato, E., \& Caprioli, D., 2012, \apj, 755, 121

\bibitem[Bret et al.(2010)]{bret10}
Bret, A., Gremillet, L., \& Dieckmann, M. E., Physics of Plasma, 17, 120501

\bibitem[Chevalier \& Raymond(1978)]{chevalier78}
Chevalier R. A., \& Raymond, J. C. 1978, \apj, 225, L27 

\bibitem[Fujita et al.(2007)]{fujita07}
Fujita, Y., Suzuki, T. K., Kudoh, T., \& Yokoyama, T. 2007, \apj, 659, L1

\bibitem[Fujita \& Ohira(2011)]{fujita11}
Fujita, Y. \& Ohira, Y. 2011, \apj, 738, 182

\bibitem[Ghavamian et al.(2007)]{ghavamian07}
Ghavamian, P., Laming, J. M., \& Rakowski, C. E. 2007, \apjl, 654, L69

\bibitem[Gnedin et al.(2000)]{gnedin00}
Gnedin, N. Y.,  Ferrara, A., \& Zweibel, E. G. 2000, \apj, 539, 505

\bibitem[Han(2017)]{han17}
Han, J. L., 2017, \araa, 55, 111

\bibitem[Hanayama et al.(2005)]{hanayama05}
Hanayama, H., Takahashi, K., Kotake, K., Oguri, M., Ichiki, K., \& Ohno, H. 2005, \apj, 633, 93

\bibitem[Harrison(1970)]{harrison70}
Harrison, E. R. 1970, \mnras 147, 279

\bibitem[Heng \& McCray(2007)]{heng07}
Heng, K., \& McCray, R. 2007, \apj, 654, 923

\bibitem[Heng(2010)]{heng10}
Heng, K. 2010, PASA, 27, 23

\bibitem[Huba \& Fedder(1993)]{huba93}
Huba, J. D., \& Fedder, J. A. 1993, Phys. Fluid B, 5, 3779

\bibitem[Keshet et al.(2009)]{kehset09}
Keshet, U., Katz, B., Spitkovsky, A., \& Waxman, E. 2009, \apjl, 693, L127

\bibitem[Krymsky (1977)]{krymsky77}
Krymsky, G. F. 1977, Dokl. Akad. Nauk SSSR, 234, 1306

\bibitem[Lesch et  al.(1989)]{lesch89}
Lesch, H, Crusius, A., Schlickeiser, R., \& Wielebinski, R. 1989, \aap, 217, 99

\bibitem[Lim \& Raga(1996)]{lim96}
Lim, A. J., \& Raga, A. C. 1996, \mnras, 280, 103

\bibitem[Miniati \& Bell(2011)]{miniati11}
Miniati, F., \& Bell, A. R. 2011, \apj, 729, 73.

\bibitem[Ohira \& Takahara(2007)]{ohira07}
Ohira, Y., \& Takahara, F. 2007, \apjl, 661, L171

\bibitem[Ohira \& Takahara(2008)]{ohira08}
Ohira, Y., \& Takahara, F. 2008, \apj, 688, 320

\bibitem[Ohira et al.(2009)]{ohira09}
Ohira, Y., Takahara, F., \& Terasawa, T. 2009, \apjl, 703, L59

\bibitem[Ohira(2012)]{ohira12}
Ohira, Y. 2012, \apj, 758, 979

\bibitem[Ohira(2013)]{ohira13}
Ohira, Y. 2013, \prl, 111, 245002

\bibitem[Ohira(2014)]{ohira14}
Ohira, Y. 2014, \mnras, 440, 514

\bibitem[Ohira(2016a)]{ohira16a}
Ohira, Y. 2016a \apj, 817, 137

\bibitem[Ohira(2016b)]{ohira16b}
Ohira, Y. 2016b, \apj, 827, 36

\bibitem[Ohira \& Murase(2019)]{ohira19}
Ohira, Y., \& Murase, K. 2019, \prd, 100, 061301(R)

\bibitem[Ohira(2020)]{ohira20}
Ohira, Y. 2020, \apjl, 896, L12

\bibitem[Ohira(2021)]{ohira21}
Ohira, Y. 2021, \apj, 911, 26

\bibitem[Pinto \& Galli(2008))]{pinto08}
Pinto, C., \& Galli, D. 2008, \aap, 484, 17

\bibitem[Pusztai et al.(2020)]{pusztai20}
Pusztai, I., Juno, J., Brandenburg, A., TenBarge, J. M., Hakim, A., Francisquez, M., \& Sundstr{\"o}n, A., 2020, \prl 124, 255102

\bibitem[Raymond et al.(2008)]{raymond08}
Raymond, J. C., Isenberg, P. A., \& Laming, J. M. 2008, \apj, 682, 408

\bibitem[Rincon et al.(2016)]{rincon16}
Rincon, F., Califano, F., Schekochihin, A. A., \& Valentini, F., 2016, PNAS, 113, 3950

\bibitem[Schultz et al.(2008)]{schultz08}
Schultz, D. R., Krstic, P. S., Lee, T. G., \& Raymond, J. C. 2008, \apj, 678, 950

\bibitem[Spitzer(1962)]{spitzer62}
Spitzer, L. 1962, Physics of Fully Ionized Gases (2nd ed.; New York: Wiley)

\bibitem[Suzuki(2002)]{suzuki02}
Suzuki, T. K. 2002, \apj, 578, 598

\bibitem[Subramanian et al.(1994)]{subramanian94}
Subramanian, K., Narasimha, N., \& Chitre, S. M. 1994, \mnras, 271, L15

\bibitem[Tomita \& Ohira(2016)]{tomita16}
Tomita, S., \& Ohira, Y. 2017, \apj, 825, 103

\bibitem[Tomita et al.(2019)]{tomita19}
Tomita, S., Ohira, Y., \& Yamazaki, R. 2019, \apj, 886, 54

\bibitem[van Adelsberg et al.(2008)]{adelsberg08}
van Adelsberg, M., Heng, K., McCray, R., \& Raymond, J. C. 2008, \apj, 689, 1089

\bibitem[Weibel (1959)]{weibel59}
Weibel, E. S. 1959, \prl, 2, 83

\bibitem[Widrow(2002)]{widrow02}
Widrow, L. M. 2002, Rev. Mod. Phys., 74, 775

\end{thebibliography}

\end{document}